\documentstyle[numreferences]{crckapb} 

\begin{opening}
\title{Effects of adsorbates on submonolayer growth}

\author{Miroslav Kotrla}
\institute{Institute of Physics
           Academy of Sciences of the Czech Republic\\
          Na Slovance 2, 182 21 Praha 8, Czech Republic}

\author{Joachim Krug}
\institute{Fachbereich Physik, Universit\"at GH Essen \\ D-45117 Essen, Germany}
\author{Pavel \v{S}milauer}
\institute{Institute of Physics, Academy of Sciences of the Czech Republic\\ 
Cukrovarnick\'a 10, 162 53 Praha 6, Czech Republic}

\end{opening}

\runningtitle{ADSORBATES IN SUBMONOLAYER GROWTH}

\begin{document}

\begin{abstract}
The effects of adsorbates on nucleation and growth of two-dimensional islands
is investigated by kinetic Monte Carlo simulations and rate equation
theory.
The variation of island morphology with adsorbate parameters
is discussed and
the temperature-dependence of island density in the case of immobile
adsorbates is studied in detail.
A set of rate equations for the description of
nucleation in the presence of predeposited
mobile and immobile adsorbates is developed.
\end{abstract}

\section{Introduction}
While a satisfactory understanding of
nucleation and growth in homoepitaxial systems is beginning
to emerge \cite{brune98}, the dramatic effects of tiny traces
of adsorbates continue to surprise researchers.
Among a wealth of recent examples, we may mention
the influence of CO on island shapes and interlayer transport on
Pt(111) \cite{kalff98}, the modification of attachment kinetics by
H preadsorbed on Si(001) \cite{smil98} and
the H-induced enhancement of Pt
self-diffusion on Pt(110) \cite{horch99}.
In this situation it seems useful to explore different generic
scenarios for adsorbate effects on submonolayer growth
within a reasonably simple, yet flexible, theoretical model.
The present paper reports on an ongoing effort
aimed in this direction.

In Section \ref{kmc} we briefly
review our earlier findings and 
present new results concerning the dependence of the 
island density and morphology on the temperature 
and the characteristics of the adsorbate 
(mobility and the strength of interaction with adatoms).
In the Sec. \ref{rate}
we sketch a rate equation theory which provides a
unified description of
two-dimensional nucleation in the presence of predeposited
mobile and immobile adsorbates, and derive analytic solutions
for some special cases.

\section{Kinetic Monte Carlo Simulations}
\label{kmc}

\subsection{Model}
We employ a recently introduced solid-on-solid growth model
with two surface species $A$ and $B$, representing the 
adatoms of the growing material and the adsorbates, respectively
\cite{kks00a,kks00b}.
The simulation starts on a flat substrate composed only of $A$-atoms.
The basic microscopic processes are deposition and migration;
desorption is not allowed.
Two deposition modes can be studied:
(i) simultaneous deposition ({\em codeposition\/}) of both species, 
with generally different  fluxes, $F_A$ and $F_B$, and
(ii) {\em predeposition\/} of a certain adsorbate coverage $\theta_B$ 
prior to growth. 

Adatoms
and adsorbates migrate with a nearest-neighbor hopping rate
$R_D = k_0 e^{-E_D/k_B T}$, where $k_0 = 10^{13}$ Hz and the
energy barrier for a particle of type $X =$ A or $B$ is given by
\begin{equation}
\label{E_D^X}
E_D^X= \sum_{Y=A,B} \left( n_0^Y E_{\rm sub}^{XY}
+ n_1^{XY} E_{\rm n} ^{XY} \right).
\end{equation}
Here $E_{\rm sub}^{XY}$ is the hopping barrier for a free $X$ atom 
on a substrate atom $Y$, $n_0^Y$ is equal to one if the substrate
atom is of type $Y$ and zero otherwise,
$n_1^{XY}$ is the number of nearest-neighbor
$X$-$Y$ pairs, and $E_{\rm n}^{XY}$
is the corresponding contribution to the barrier (symmetric in $X$ and $Y$).
There are no lateral interactions between adsorbate atoms 
($E_{\rm n}^{BB}=0$). The diffusion rate of a free adatom
is $D = k_0 e^{-E_{\rm sub}^{AA}/k_B T}$.

In previous work \cite{kks00a}, we have shown that
in addition a process of place exchange between an adsorbate
and an adatom with at most one lateral bond is necessary to achieve
decorated island edges. This occurs at a rate
$k_0 e^{-E_{\rm ex}/k_B T}$. 
When the barrier  $E_{\rm ex}$
is sufficiently low, then
adsorbates are floating on the edges of growing islands.

Adsorbates can be mobile or immobile and strongly or weakly interacting 
with the adatoms. 
Their mobility is controlled by the strength of interaction
of adsorbates with the substrate which is described by the energy 
barrier $E_{\rm sub}^{BA}$. The interaction with adatoms is determined 
by the energy barrier $E_{\rm n}^{AB}$.

\subsection{Summary of previous results}
\label{Summary}
We have investigated previously the case of floating mobile adsorbates
in the situation when the concentration of adsorbates is
comparable with that of the growing material \cite{kks00b}. 
We studied the dependence of the island density $N$ on the
flux $F_A$ and the coverage $\theta_A$ for both codeposition and predeposition.
We found that the adsorbates  strongly increase the island density 
without appreciably changing its power-law dependence on flux,
the increase being stronger for larger $E_{\rm n}^{AB}$.
The increase was only slightly higher for predeposition than
for codeposition. 

We also observed a stronger coverage dependence of the island density
in comparison with homoepitaxy.
This was interpreted as 
a delay of the saturation regime, where
island density becomes independent of coverage. 
The coverage dependence is again more 
pronounced for strongly interacting adsorbates,
but it is weaker in the case of predeposition than for codeposition.
A further noteworthy feature is the
much higher density of free adatoms
than in homoepitaxy, which shows an
intriguing, oscillatory coverage dependence
in the case of codeposition \cite{kks00b}.

Many of these simulation results could be qualitatively
explained in terms of a simple rate equation theory,
which rests on the assumption that 
the adsorbates affect the growth process only by slowing down the
diffusion of adatoms. If, in addition, the adsorbates are treated
as immobile traps (which may be justified at least for 
sufficiently high adsorbate coverage $\theta_B$, see Section
\ref{RateMob}), the adatom migration is
described by an effective diffusion coefficient
\begin{equation}
\label{Dbar}
\overline D(\theta_B) = \frac{D}{1 - \theta_B + 
\theta_B e^{E_{\rm n}^{AB}/k_B T}}.
\end{equation}

Recently, we have considered a situation when only a low concentration of 
predeposited adsorbates is present on the surface \cite{kks00c}. 
We investigated the dependence of island density on flux
and on the concentration of adsorbates $\theta_B$. 
Even concentrations as low as 
$\theta_B=0.002$~ML can lead to a severalfold increase of the
island density.
However, the increase significantly depends on the
mobility of the adsorbates. In the case of essentially 
immobile adsorbates ($E_{\rm sub}^{BA}=5$ eV),
a new feature in the flux dependence of the island
density appears. Instead of a single power law relationship,
there is a plateau where $N \approx \theta_B$,
reflecting the dominance of heterogeneous nucleation  
in a flux interval  $F_1 > F > F_2$. 
Scaling arguments yield the estimates 
$ F_1 \approx D \theta_B^2$,
$F_2 \approx \theta_B D e^{-E_{\rm n}^{AB}/k_B T}$
for the plateau boundaries in terms of impurity
coverage and strength. For sufficiently small 
$\theta_B$ these were confirmed by scaling plots
of $N/\theta_B$ vs. $F/F_1$ and $F/F_2$ \cite{kks00c}.

In this work we further extend the above results.
In the following subsection, we discuss effects of 
adsorbates on the island morphology, and in the subsequent subsection
we present results on the temperature dependence of the island density.
The energy barriers 
$E_{\rm sub}^{AA}=0.8$ eV, $E_{\rm sub}^{AB}=0.1$ eV, 
$E_{\rm sub}^{BB}=0.1$ eV, and
$E_{\rm n}^{AA}=0.3$ eV were fixed in the following. 
Other energy barriers as well as the ratio $D/F$ were varied.

\subsection{Island morphology}

The island shape in homoepitaxy depends on the rates of the
various kinetic processes involved. 
The size of islands varies with the ratio $D/F$.
To be specific, we set in this subsection the temperature  to $T=500 $~K.
Then, the island shapes in homoepitaxy with the parameters 
mentioned above in Section \ref{Summary} are
regular (approximately rectangular, close to square shape,
nondendritic) for fluxes in 
the interval $F = 0.00025 - 0.64$ ML/s.
The shapes become irregular 
for higher fluxes. In order to evaluate the effect of adsorbates,
we performed several sets of simulation with fixed flux $F = 0.001$ ML/s
and varying adsorbate properties.

Let us consider first the case of floating adsorbates.
It is useful to distinguish two situations: (i) well decorated island
and (ii) poorly decorated islands.
In the former case, the majority of edge sites is decorated by an adsorbate.
We found that the shape is very regular, only the islands
are smaller; in fact it can be shown that the decoration
{\em enhances} edge diffusion and thus makes the island edges
smoother \cite{kks00b}.
The shape is not very sensitive to the mobility of adsorbates. It only
has to be sufficiently high so that adsorbates can reach the island edges.
The exchange of adatoms approaching the step edge
and the adsorbates decorating the step edge
is the dominant mechanism.

If the islands are poorly decorated (the number of adsorbates
attached to the island edges 
is considerably less than the number of perimeter sites),
then, the shape changes to irregular and regularity is restored 
again only for very small $\theta_B$. 
Rather than promoting edge diffusion, in this regime the 
adsorbates bound on the island edges act as sinks for adatoms 
provided that the adatom-adsorbate interaction (given by $E_{\rm n}^{AB}$)
is sufficiently strong.
Another factor affecting the island shape in this situation is
the change of the mobility 
of adsorbates (given by  $E_{\rm sub}^{BA}$). 
Examples of the different island mophologies for various values of parameters 
$E_{\rm n}^{AB}$ and  $E_{\rm sub}^{BA}$
are shown in Fig. 1.
We can see that the increase of both energy barriers leads to more irregular
island shapes and also to an increase of island density.

\newpage
\begin{figure}[t]
\centering
\label{fig:conf}
\vspace*{159mm}
\includegraphics{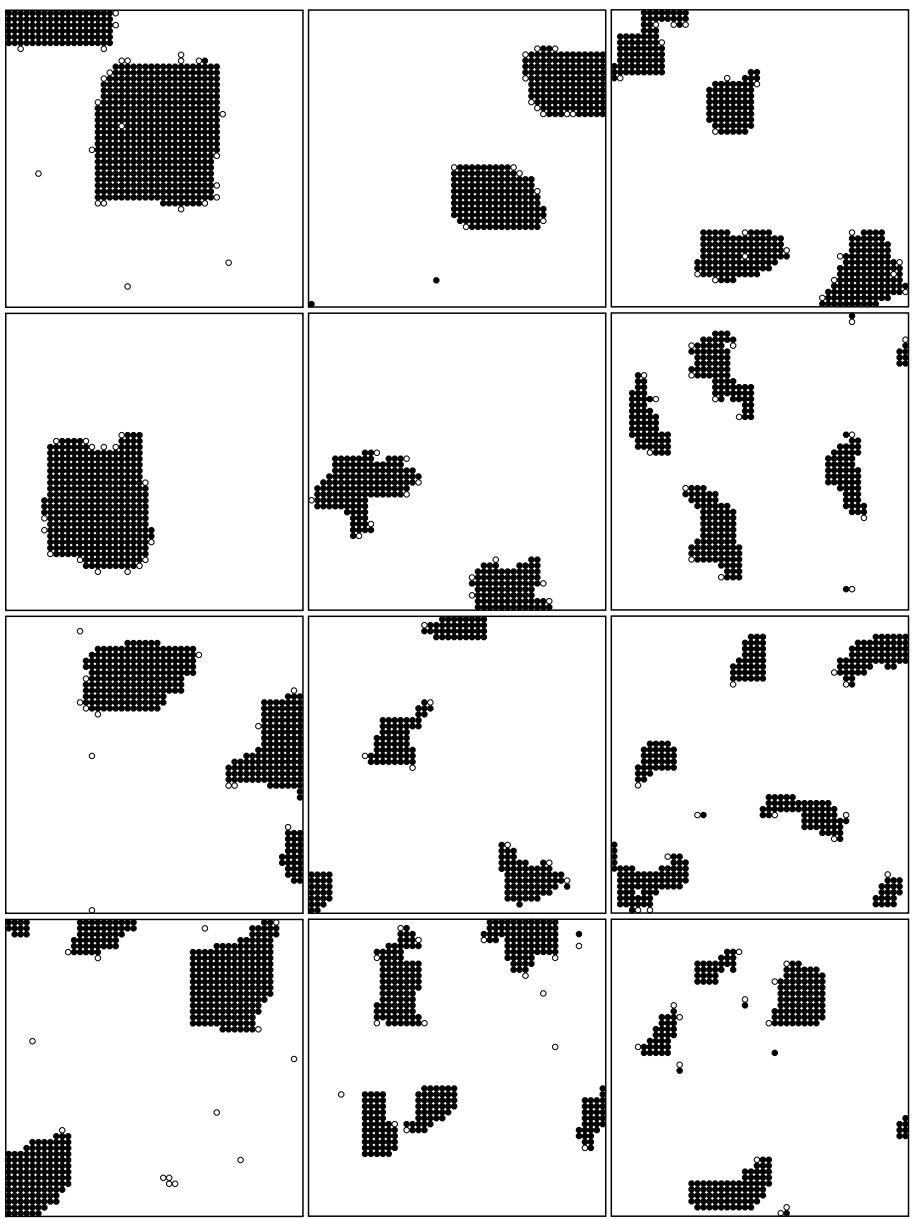}
\caption{
Examples of configurations obtained
with flux $F_A = 0.001$~ML/s
at a  coverage $\theta_A = 0.1$~ML after predeposition
 $\theta_B = 0.005$ ML of  adsorbate coverage,
$E_{\rm n}^{AA} = 0.3$~eV,   $E_{\rm ex} = 1$~eV
and different energy barriers 
$E_{\rm n}^{AB} = 0.2$~eV (left column),
$E_{\rm n}^{AB} = 0.4$~eV (middle column),
$E_{\rm n}^{AB} = 0.6$~eV (right column),
$E_{\rm sub}^{BA}$: from top to bottom
  $E_{\rm sub}^{BA} = 0.7$~eV,
  $E_{\rm sub}^{BA} = 1.0$~eV,
  $E_{\rm sub}^{BA} = 1.2$~eV,
  $E_{\rm sub}^{BA} = 1.5$~eV.
We show only 50 $\times$ 50 sections of larger simulation
boxes.
}
\end{figure}

\newpage
\noindent
The irregularity is not so apparent for weak adsorbate-adatom interaction
(left column) or high mobility of adsorbates (top row).
The immobile strongly interacting adsorbate catches more diffusing adatoms
and at the same time it is floating on the  
growing island edge. Hence, the island grows faster 
in the neighborhood of such adsorbates.
This kind of adsorbates also
block the motion of adatoms along the edge and thus 
suppresses  the smoothening of the island edges.

Let us discuss briefly the island morphology for non-floating adsorbates.
The mechanism of floating becomes less relevant for adsorbates 
with low mobility and does not act at all in the limit
of immobile adsorbates.
Non-floating adsorbates are incorporated inside a growing island
and beyond a certain island size have no effect on growth.
In this context, we want to note that
surface defects can be treated in our approach as immobile adsorbates.

\subsection{Temperature dependent island density}

While in our simulations we have so far focused on the flux
dependence of the island density [5-7]
in experiments  it is most often measured as a function of 
temperature  at fixed flux.
Therefore, we complement our previous results by the calculation
of the temperature dependence.
We restrict ourselves 
here to the case of predeposition of immobile (non-floating)
adsorbates. In Fig.2 we show data
for several adsorbate concentrations. 
We find, similar to the flux dependence \cite{kks00c},
a plateau interval
$T_1 < T < T_2$, where $N$ is almost independent of $T$.
For low temperatures, $T < T_1$, homogeneous nucleation is more important
than heterogeneous nucleation by adsorbates, because the adatoms
are not sufficiently mobile to reach the adsorbates.
In the plateau region nucleation is predominantly heterogeneous. 
For high temperatures, $T > T_2$, adatoms start to detach from adsorbates 
and homogeneous nucleation is again important. This kind of behavior
is observed experimentally in defect nucleation on oxide and halide
surfaces \cite{ven00} and in growth on surfaces with strain-relief
patterns, which act as an ordered array of defects
\cite{bru98}.  

For a quantitative estimate of the transition temperatures
$T_1$ and $T_2$, we recall the
three relevant time scales in the problem \cite{kks00c}:
(i) The {\it diffusion time} $\tau_D \approx 1/D \theta_B$,
which is the time required for an atom to explore the ``capture zone''
of area $1/\theta_B$ associated with a single impurity, and hence
to get trapped at the impurity. (ii) The {\it deposition time}
$\tau_F \approx \theta_B/F$, which is the time between subsequent
arrivals of adatoms within the capture zone. (iii) The trapping
time $\tau \approx (1/D) e^{E_{\rm n}^{AB}/k_B T}$ at the impurity.
The plateau regime is characterized by $\tau > \tau_F > \tau_D$,
which yields the expressions
\begin{equation}
\label{T1T2}
T_1 = \frac{E_{\rm sub}^{AA}}{k_B \ln(k_0 \theta_B^2/F)}, \;\;\;
T_2 = \frac{E_{\rm sub}^{AA} + E_{\rm n}^{AB}}{k_B \ln(k_0 \theta_B/F)}.
\end{equation} 
Using $E_{\rm sub}^{AA} = 0.8$ eV,
$F=0.001$ ML/s and $\theta_B =0.001$ ML, we get $T_1 = $  403 K,
while $T_2 =$ 465 K for $E_{\rm n}^{AB} = 0.4$ eV and
$T_2 =$ 542 K for $E_{\rm n}^{AB} = 0.6$ eV,
in good agreement with the simulations.
From the expression for $T_2$, we directly see how the width of the plateau
grows with increasing  $E_{\rm n}^{AB}$.

\begin{figure}[ht]
\centering
\vspace*{80mm}
\includegraphics{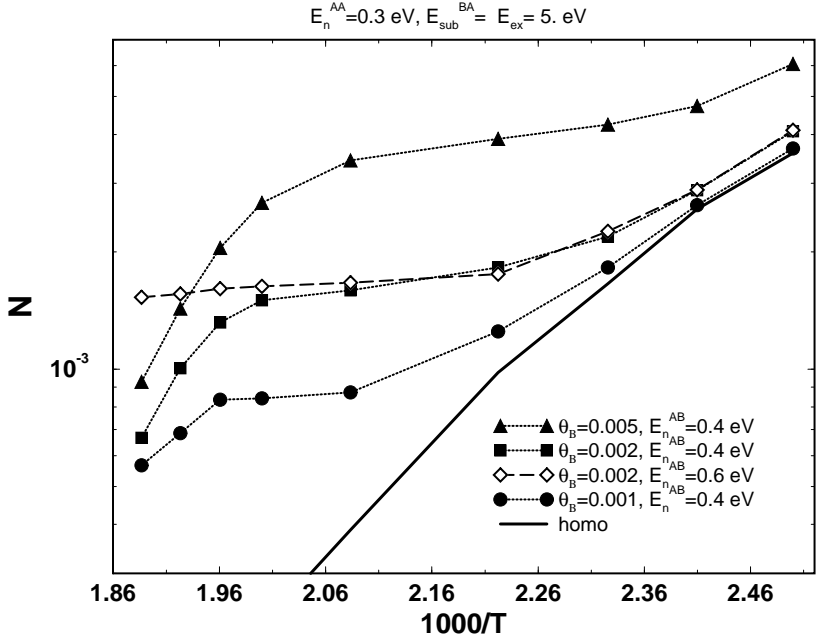}
\caption{
Averaged island density as a function of the inverse temperature
for different concentration  $\theta_B$ of predeposited impurities
and different  energy of interaction between adsorbates and adatoms:
$E_{\rm n}^{AB} = 0.4$~eV - filled symbols,
$E_{\rm n}^{AB} = 0.6$~eV - open symbols.
The adatom interaction energy
$E_{\rm n}^{AA} = 0.3$~eV and the energy barriers
$E_{\rm sub}^{BA} = E_{\rm ex} = 5$~eV
are fixed. Deposition flux is $F_A = 0.001$~ML/s,
the coverage of adatoms is $\theta_A = 0.1$~ML.
The behavior in the absence of impurities
(homoepitaxy, $\theta_B = 0$~ML) is shown for comparison.
}
\label{fig:temper}
\end{figure}

\section{Rate Equations}
\label{rate}
In this section we develop a set of rate equations for the description of
two-dimensional nucleation in the presence of mobile and immobile adsorbates.
Throughout, the impurities will be assumed to have been predeposited 
at a coverage
$\theta_B$, though codeposition is easily incorporated along the lines of
our earlier work \cite{kks00b}.
To keep matters transparent, we also restrict ourselves to the case of
irreversible aggregation (critical nucleus size $i^\ast = 1$). Detachment of impurities
from adatom islands will, however, be included, since this process plays a crucial role
at low deposition fluxes.

In addition to the adatom density $n$ and the island density $N$, a minimal description
of nucleation in the presence of adsorbates turns out to require the monitoring
of two additional quantities \cite{ven00,ven97}:
The density of free impurities $n_0$, and the
density $n_c$ of critical clusters bound at impurities; in the present case $i^\ast = 1$
the latter is simply the density of bound adatom-impurity pairs. Taking into account
all relevant processes, 
one arrives at the following set of evolution equations:
\begin{equation}
\label{raten}
dn/dt = F - Dn(2n + N) - D n(n_0 + n_c) + n_c/\tau ,
\end{equation}
\begin{equation}
\label{rateN}
dN/dt = D n(n + n_c) ,
\end{equation}
\begin{equation}
\label{raten0}
dn_0/dt = - D n n_0 - \tilde D n_0 N +  n_c/\tau  + N/\tau' ,
\end{equation}
\begin{equation}
\label{ratenc}
dn_c/dt = D n (n_0 - n_c) - n_c/\tau .
\end{equation}
Here $1/\tau \approx k_0 e^{-(E_{\rm sub}^{AA} + E_{\rm n}^{AB})/k_B T}$ denotes the rate of
detachment of an adatom from an impurity,
$\tilde D = k_0 e^{-E_{\rm sub}^{BA}/k_B T}$ is the impurity diffusion constant, and
$1/\tau' \approx k_0 e^{-(E_{\rm sub}^{BA} + E_{\rm n}^{AB})/k_B T}$ is the detachment rate
of impurities from islands. In the expression for $\tau$, 
we have assumed that the impurities are
(much) less mobile than the adatoms, so that the relative mobility is dominated by the
adatom mobility.

Most of the impurity terms in Eqs.(\ref{raten}-\ref{ratenc}) should be self-explanatory.
We merely point out the term $n_c/\tau$ in (\ref{raten},\ref{raten0},\ref{ratenc}) which
describes the dissociation of adatom-impurity pairs, and the term $N/\tau'$ in (\ref{ratenc})
which describes detachment of impurities from the islands. 
Strictly speaking, all reaction terms
should be adorned with dimensionless capture coefficients. 
However, since our objective here is
a qualitative, rather than a quantitative description, all these coefficients have been set to unity.

\subsection{Dimensionless formulation}

Following Tang \cite{tang93} we now rescale the adatom and island densities by $\sqrt{F/D}$ and
time by $1/\sqrt{DF}$. In addition, 
the impurity densities $n_0$ and $n_c$ will be rescaled by
the initial impurity coverage $\theta_B$. Marking dimensionless quantities with a hat, we obtain
the rescaled evolution equations
\begin{equation}
\label{ratenhat}
d \hat n/d \hat t = 1 - \hat n(2 \hat n + \hat N) - a \hat n(\hat n_0 + \hat n_c) + b \hat n_c ,
\end{equation}
\begin{equation}
\label{rateNhat}
d \hat N/d \hat t = \hat n(\hat n + a \hat n_c) ,
\end{equation}
\begin{equation}
\label{raten0hat}
d\hat n_0/d \hat t = - \hat n \hat n_0 - c \hat n_0 \hat N +  (b/a) \hat n_c  + (bc/a^2) \hat N ,
\end{equation}
\begin{equation}
\label{ratenchat}
d\hat n_c/d \hat t = \hat n (\hat n_0 -\hat n_c) -(b/a) \hat n_c ,
\end{equation}
where $a = \theta_B \sqrt{D/F}$, $b = \theta_B /F \tau$ and
$c = \tilde D/D$. The initial conditions are
$\hat n(0) = \hat N(0) = \hat n_c(0) = 0$, $\hat n_0(0) = 1$.

The parameters $a$ and $b$ may be written in terms
of the characteristic fluxes, $F_1$ and $F_2$, which limit the plateau regime in the case of immobile impurities, as 
$a = \sqrt{F_1/F}$ and $b = F_2/F$.
For impurities to be at all relevant, 
it is necessary that $F_1/F_2 = D \theta_B \tau = a^2/b \gg 1$.
By retaining only the most important terms,
we can analytically extract the behavior predicted by Eqs.(\ref{ratenhat}-\ref{ratenchat}) in simple cases.
So far this has been achieved only for immobile impurities ($c=0$).

\subsection{Immobile impurities without detachment}

Here we consider the plateau regime, where
$a \gg 1$ and $b \ll 1$. Numerical integration of (\ref{ratenhat}-\ref{ratenchat}) with
$b = c = 0$ and $a \gg 1$ shows that the adatom density is time independent, taking the value
$n = F/D \theta_B$ to high accuracy. For early times, 
this reflects the balance between deposition
and capture at impurities, while at late times the capture at islands dominates; since $N = \theta_B$ at
late times, the resulting adatom density is the same in both regimes. 
Setting $n \equiv F/D \theta_B$,
the remaining three rate equations become linear and are readily integrated, with the result
\begin{equation}
\label{Nlin}
N = \theta_B (1 - (1 + \theta/\theta_B)e^{-\theta/\theta_B}) ,
\end{equation}
\begin{equation}
\label{n0lin}
n_0 = \theta_B e^{-\theta/\theta_B} ,
\end{equation}
\begin{equation}
\label{nclin}
n_c = \theta e^{-\theta/\theta_B}.
\end{equation}
Both impurity species decay exponentially at late times.

\subsection{Immobile impurities with detachment}

The regime $F \ll F_2$ is detachment-dominated, in the sense that an adatom
typically visits many impurities before being incorporated in an island. The primary effect of the
impurities is then to slow down the diffusion by temporarily trapping the adatoms \cite{kks00b}.
To see how this emerges from the rate equations, we infer from numerical integration that, in
the relevant late time regime, an equilibrium is established between attachment and detachment
of adatoms at free impurities. This implies that $D n n_0 \approx n_c/\tau$.
Equation (\ref{raten0}) then yields
$dn_0/dt \approx 0$, so that the impurity concentration essentially retains its initial value
$n_0 = \theta_B$, and $n_c \approx (D \theta_B \tau) n$. In this way, 
the impurity rate equations
(\ref{raten0},\ref{ratenc}) have effectively been eliminated, and we are left with a pair
of modified equations for $n$ and $N$. Taking into account that $D \theta_B \tau = F_1/F_2 \gg 1$,
and therefore $n_c \gg n$, they read
\begin{equation}
\label{neff}
dn/dt = F - Dn[(D \theta_B \tau) n + N] ,
\end{equation}
\begin{equation}
\label{Neff}
dN/dt = (D \theta_B \tau) D n^2.
\end{equation}
Standard analysis \cite{tang93} shows that the asymptotic island density is of the order
$N \sim (F \tau \theta_B)^{1/3}$, which can be brought into the familiar form
$N \sim (F/\bar D)^{1/3}$ by identifying the effective diffusion constant 
as $\bar D = 1/(\tau \theta_B)$. This is 
precisely the strong impurity limit of (\ref{Dbar}).
\begin{figure}[ht]
\centering
\label{fig:rate}
\vspace*{75mm}
\includegraphics{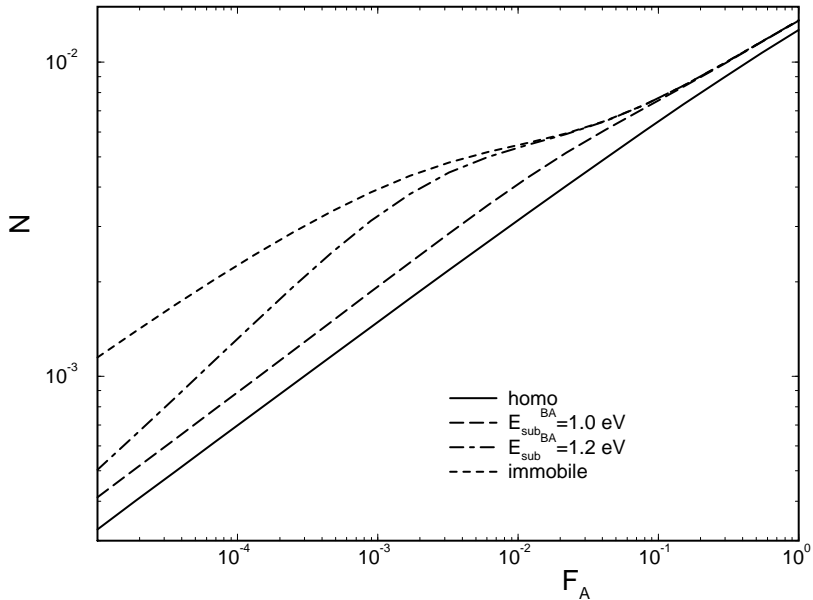}
\caption{Island density as a function of deposition flux,
obtained by numerical solution of the
rate equations (\ref{raten}-\ref{ratenc}) for different choices of
the impurity diffusion barrier.
The adatom coverage is $\theta = 0.1$ ML and the impurity coverage $\theta_B = 0.005$ ML.}
\end{figure}

\subsection{The effect of impurity mobility}
\label{RateMob}

The second term on the right-hand side of Equation (\ref{raten0}) describes the depletion
of free impurities due to capture at stable islands. This is partly compensated
by the detachment of impurities from islands described by the last term, 
but nevertheless,
the net effect is expected to be a decrease of the island density, because the depletion
reduces the ability of mobile impurities to act as nucleation centers.

In the absence of
an analytic study of the full set of rate equations, in Fig. 3 we provide
some sample results obtained by numerical integration. The adatom
diffusion constant $D$ and the detachment time $\tau$ were chosen in accordance with the
parameters employed in the KMC simulations. Data for three different choices of the impurity
diffusion barrier $E_{\rm sub}^{BA}$ are shown, and the behavior in the absence of impurities
is included for reference. The case of low impurity mobility ($E_{\rm sub}^{BA} = 1.2$ eV, corresponding
to $\tilde D/D \approx 10^{-4}$) is particularly noteworthy, as it indicates an intermediate scaling
regime where the island density scaling exponent, defined by $N \sim F^\chi$, is larger than the homoepitaxial
value $\chi = 1/3$ (a fit yields $\chi \approx 0.42$). 
Further exploration of this phenomenon will be presented
elsewhere.

\paragraph{Acknowledgments.} J.K. would like to thank CAMP and the Department of Physics at
DTU for their kind hospitality while this paper was prepared. Support by the COST project P3.130 and
by Volkswagenstiftung are gratefully acknowledged.


\end{document}